\documentclass[]{spie}  

\usepackage{color}
\usepackage{graphicx}
\usepackage{subfigure}
\usepackage{bm}
\usepackage{amsmath}
\usepackage{amssymb}
\usepackage{cite}
\usepackage{stfloats}
\usepackage{bm}
\usepackage{mathrsfs}

\usepackage{subfigure}
\usepackage{amsmath}
\usepackage{amssymb}
\usepackage{stfloats}

\title{Three-dimensional optical diffraction tomographic microscopy with optimal frequency combination with partially coherent illuminations}

\author[]{Jiaji Li$^{1,2,3,4}$}
\author[]{Qian Chen$^{1,2}$}
\author[]{Jiasong Sun$^{1,2,3}$}
\author[]{Jialin Zhang$^{1,2,3}$}
\author[]{Junyi Ding$^{1,2,3}$}
\author[]{Chao Zuo$^{1,2,3,*}$}

\affil[]{$^1$School of Electronic and Optical Engineering, Nanjing University of Science and Technology, No. 200 Xiaolingwei Street, Nanjing, Jiangsu Province 210094, China\\
$^2$Jiangsu Key Laboratory of Spectral Imaging \& Intelligent Sense, Nanjing University of Science and Technology, Nanjing, Jiangsu Province 210094, China\\
$^3$Smart Computational Imaging Laboratory (SCILab), Nanjing University of Science and Technology, Nanjing, Jiangsu Province 210094, China\\
$^4$jiajili@njust.edu.cn\\}

\authorinfo{Further author information: (Send correspondence to C. Zuo)\\E-mail: surpasszuo@163.com}

\begin{document}
\maketitle

\begin{abstract}
We demonstrate a three-dimensional (3D) optical diffraction tomographic technique with optimal frequency combination (OFC-ODT) for the 3D quantitative phase imaging of unlabeled specimens. Three sets of through-focus intensity images are captured under an annular aperture and two circular apertures with different coherence parameters. The 3D phase optical transfer functions (POTF) corresponding to different illumination apertures are combined to obtain an optimally synthesized frequency response, achieving high-quality, low-noise 3D reconstructions with imaging resolution up to the incoherent diffraction limit. Besides, the 3D imaging performance of annular illumination is explored and the expression of 3D POTF for arbitrary illumination pupils is derived and analyzed. It is shown that the phase-contrast washout effect in high-NA circular apertures can be effectively addressed by introducing a complementary annular aperture, which strongly boosts the phase contrast and improves the practical imaging resolution. To test the feasibility of the proposed OFC-ODT technique, the 3D refractive index reconstruction results based on a simulated 3D resolution target and experimental investigations of micro polystyrene bead and unstained biological samples are presented. Due to its capability of high-resolution 3D phase imaging as well as the compatibility with widely available commercial microscope, the OFC-ODT is expected to find versatile applications in biological and biomedical research.
\end{abstract}

\section{Introduction}
The refractive index (RI) of biological cells and tissues contain important biophysical information about shapes, sizes, volumes and dry mass, and these characteristics are crucial for the morphological detection and diagnosis of disease. There are several qualitative and quantitative methods have been proposed to provide reliable rendering about phase contrast and quantitative phase change introduced by the heterogeneous RI distribution within the label-free samples. Zernike phase contrast (PhC) microscopy \cite{PCM} and differential interference contrast (DIC) microscopy \cite{DIC} can effectively visualize nearly transparent biological samples by converting the phase into intensity, which have been widely used in cell biology studies. Alternatively, quantitative phase imaging (QPI) microscopy provides the quantitative data interpretation and phase reconstruction, and this imaging technique can be realized by the interferometric and non-interferometric approaches. The conventional off-axis digital holographic microscopy (DHM) \cite{DMH1,DMH2,DMH3} has been developed to measure the total phase delay quantitatively, and some other interferometric methods based on common path geometries have been also proposed to improve the imaging quality and spatial resolution of the phase measurement by using spatial light modulator (SLM) or white-light sources \cite{GP1,GP2}. In addition, there are also numerous variants of non-interferometric phase retrieval approaches, like transport of intensity equation (TIE) \cite{TIE1,TIE2,TIE3,TIE4,TIE5,TIE6} and differential phase contrast (DPC) \cite{2D_DPC}, which provide promising QPI results under coherent illumination and partially coherent illumination.

When the thickness of object is larger than the depth of field (DOF) of microscopy, a single two-dimensional (2D) integral phase image is insufficient for the characterization of heterogeneous objects, and the detailed volumetric information inside the sample is inaccessible. Three-dimensional (3D) RI distribution indeed enables the intracellular observation of biological samples, which can be recovered based on different tomographic approaches. The off-axis and common-path DHM are applied to the implementation of 3D label-free RI imaging with computerized tomography or optical diffraction tomography (ODT) based on the Fourier diffraction theorem \cite{CT1,ODT_Ori1}. In coherent ODT (C-ODT), the incident angle of coherent beam is changed by the rotating mirrors \cite{ODT_Coh1,ODT_Coh2,ODT_Coh3,ODT_Coh4}, liquid crystals SLM \cite{TOMO_SLM}, or digital micro-mirror device (DMD) \cite{TOMO_DMD} or light-emitting diode (LED) array \cite{TOMO_LED1,TOMO_LED2}. The maximum coverage of the illumination angles is determined by the NA of the condenser lens and the incomplete range of incident angles due to the finite NA leads to a missing cone problem. Although the mechanical rotation of the object enables a more complete angular coverage \cite{ODT_NonFiber}, the radial run-out of rotation inevitably makes the sample unstable, and this C-ODT method is more suitable for certain solid non-biological objects such as optical fibers \cite{ODT_Fiber1,ODT_Fiber2}.

On a different note, ODT can also be implemented based on a conventional bright-field microscope with partially coherent illuminations \cite{PC_ODTOri}. In partially coherent ODT (PC-ODT), the 3D RI distribution can be directly recovered from a through-focus intensity stack based on the 3D phase optical transfer function (POTF) of the imaging system. For a condenser aperture with a determined NA, the spatial frequency components of the object scattering potential transmitted by the PC-ODT are the same as the C-ODT under various angles of illumination within the maximum NA. Thus, the PC-ODT method is very suitable for 3D quantitative phase imaging based on a traditional bright-field microscope. Many interesting PC-ODT approaches have been developed very recently. M. Chen \emph{et al.} \cite{3D_DPC} extended the 2D DPC quantitative imaging into 3D RI measurement by introducing an additional axial scanning process. The 3D RI reconstruction is achieved by a 3D FFT-based deconvolution, without an intermediate 2D phase retrieval step. Y. Bao and T. Gaylord \cite{Bao_3DOTF} extended Streibl's OTF theory and derived the analytical expression of 3D POTF for the non-paraxial case. T. Alieva \emph{et al.} \cite{Soto1,Soto2,Soto3} adopted the non-paraxial POTF model for 3D phase deconvolution and obtained excellent 3D RI reconstruction results of biological samples recently. However, as pointed in \cite{Soto1}, the phase contrast progressively vanishes as the illumination NA approaches the objective NA, suggesting that the phase information can hardly be transferred into intensity when the illumination NA is large. In the case of partially coherent illumination, the maximum achievable lateral resolution is determined by the sum of the objective NA and the illumination NA, where the ratio of illumination NA to objective NA ($N{A_{ill}}/N{A_{obj}}$) is so-called coherence parameter \cite{PC_ODTOri}. In other words, the intensity image gives no phase contrast under incoherent illumination when the coherence parameter equals 1 ($N{A_{ill}} = N{A_{obj}}$ actually). Despite the doubled lateral resolution, the amplitude of the 3D POTF is significantly attenuated and the signal-to-noise ratio (SNR) of intensity stack is too poor to recover the 3D phase distribution. So, there is an inherent tradeoff between phase contrast and imaging resolution in the PC-ODT, which prevents the maximum possible resolution (2$N{A_{obj}}$) for 3D phase imaging. Note that the similar phenomenon has previously been observed in defocus-based 2D QPI techniques like the TIE \cite{TIE1,TIE2,TIE5}.

In this work, we demonstrate a 3D optical diffraction tomographic technique with optimal frequency combination (OFC-ODT) for the 3D quantitative phase imaging of unlabeled specimens based on a commercial inverted microscope. Three through-focus intensity stacks are captured under three illumination sources including an annular aperture and two circular apertures with different coherent parameters. The optimal frequency components of 3D POTF corresponding to different illumination apertures are combined together by utilizing linear least-squares method to obtain a more accurate 3D reconstruction result with imaging resolution up to the incoherent diffraction limit. Although many previous works have provided complete theories about 3D partially coherent imaging \cite{PC_ODTOri,Bao_3DOTF} and demonstrated promising 3D RI experimental results \cite{Soto1,Soto2,Soto3} based on circular illumination aperture, a straightforward and comprehensible numerical expression of 3D POTF for arbitrary illumination apertures has not been developed yet. The novelty of this work is to propose a tomographic technique with OFC by invoking the annular source and re-derive the expression of 3D POTF for arbitrary illumination apertures, especially for annular illumination aperture. The annular aperture enhances the amplitude of the POTF on the Ewald sphere for both low and high frequencies in Fourier space. Not only the lateral resolution can be extended to 2NA of objective, but also the contrast of captured intensity image is high enough in our method even when the condenser aperture is fully opened. Furthermore, the 3D RI reconstruction results based on a simulated 3D resolution target and experimental analysis of micro polystyrene bead for different apertures validate the improvement of resolution and the accuracy of this method. Finally, experimental investigations and 3D rendering of unstained biological samples are presented, demonstrating the technique's practical capabilities and versatility for a wide range of applications in biomedical community.

\section{Re-derivation of 3D OTF for arbitrary illumination source}

The main purpose of ODT is to reconstruct the 3D RI distribution of the specimen, but this information is included in the optical scattering potential in heterogeneous medium. The scattering potential of object is defined by the function $V\left( {\bf{r}} \right) = k_0^2\left[ {{n^2}\left( {\bf{r}} \right) - n_m^2} \right]$, where $k_0$ is the wave number ${{2\pi } \mathord{\left/ {\vphantom {{2\pi } {{\lambda _0}}}} \right.\kern-\nulldelimiterspace} {{\lambda _0}}}$ with ${\lambda _0}$ being the wavelength in free-space, $n(\bf{r})$ and $n_m$ are the RI of specimen and its surrounding medium, correspondingly. The complex RI $n\left( {\bf{r}} \right)$, which equals ${n_p}\left( {\bf{r}} \right) + i \cdot {n_a}\left( {\bf{r}} \right)$, contains the sample's RI $n_p$ and the absorptivity $n_a$. While the sample only modulates the phase of transmitted field, the complex RI and scattering potential functions are always real.

In bright-field transmission microscope, N. Streibl \cite{PC_ODTOri} and Y. Bao \emph{et al.} \cite{Bao_3DOTF} presented the analytical form of 3D OTF for partially coherent illumination with circular aperture in the paraxial and non-paraxial regime, respectively. Before the derivation of 3D OTF formula for arbitrary illumination source, let us review some previous works about the 3D image formation in a partially coherent microscope. 3D image formation under partially coherent illumination can be described as a 3D convolution between the light source and objective pupil by invoking the Born approximation (weak scattering object) \cite{PC_ODTOri}. The measured intensity of image stack is expressed as a linear superposition of the real and imaginary parts of the object scattering potential convolved with the corresponding point spread functions (PSFs) ${H_P}\left({\bf{r}}\right)$ and ${H_A}\left({\bf{r}}\right)$:
\begin{equation}\label{Eq1}
I\left( {\bf{r}} \right) = B + \Phi\left( {\bf{r}} \right) \otimes {H_P}\left( {\bf{r}} \right) + A\left( {\bf{r}} \right) \otimes {H_A}\left( {\bf{r}} \right)
\end{equation}
where $B$ is the background intensity and can be understood as the un-scattered light or transmitted light, $\Phi\left( {\bf{r}} \right)$ and $A\left( {\bf{r}} \right)$ are the respective real and imaginary parts of object scattering potential function. Implementing the 3D FFT to above Eq. (\ref{Eq1}), the 3D spectrum of intensity images is given by the sum of delta function and the product of scattering potential spectrum with corresponding transfer function in Fourier space:
\begin{equation}\label{Eq2}
\widetilde I\left( {\bm{\zeta }} \right) = B\delta \left( {\bm{\zeta }} \right) + \widetilde \Phi\left( {\bm{\zeta }} \right){T_P}\left( {\bm{\zeta }} \right) + \widetilde A\left( {\bm{\zeta }} \right){T_A}\left( {\bm{\zeta }} \right)
\end{equation}
where $\widetilde \Phi\left( {\bm{\zeta }} \right)$, $\widetilde A\left( {\bm{\zeta }} \right)$, ${T_P}$ and ${T_A}$ are the 3D phase spectrum, absorption spectrum, POTF and amplitude optical transfer function (AOTF), respectively.

\begin{figure}[!t]
    \centering
    \includegraphics[width=13cm]{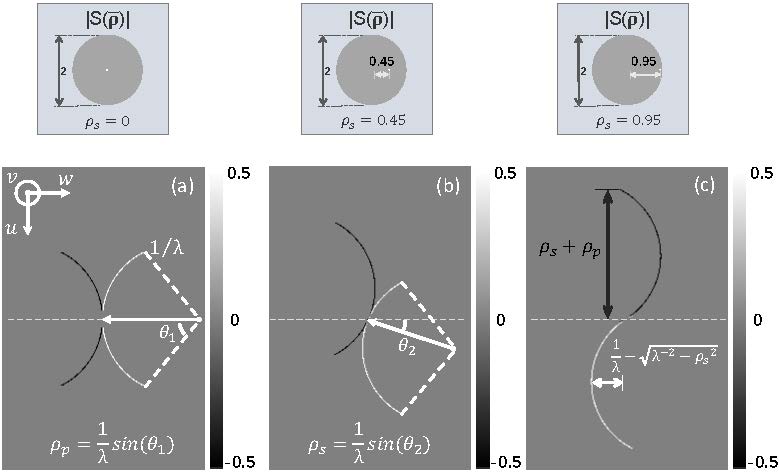}
    \caption{2D sections of 3D POTF for oblique coherent source with different normalized illumination NA. The radius of Ewald sphere is $1/\lambda$, and the normalized objective NA $\rho_p$ and illumination NA $\rho_s$ are $\sin \left( {{\theta _1}} \right)/\lambda$ and $\sin \left( {{\theta _2}} \right)/\lambda$, respectively. The lateral and axial resolution limits are $\rho_s + \rho_p$ and $1/\lambda - \sqrt {{\lambda ^{ - 2}} - \rho _s^2}$, respectively.}
    \label{Fig1}
\end{figure}

Here, we eliminate the complicated derivation steps of 3D OTF in \cite{PC_ODTOri} and give the generally existing 3D POTF of microscope in non-paraxial regime \cite{Bao_3DOTF,Annu_ODT}:
\begin{equation}\label{Eq3}
\begin{aligned}
{T_P}\left( {{\bm{\rho }},w} \right) = \frac{{j\lambda }}{{4\pi }}\iint{} & P\left( {{\bm{\rho '}} + \frac{1}{2}{\bm{\rho }}} \right){P^*}\left( {{\bm{\rho '}} - \frac{1}{2}{\bm{\rho }}} \right) {\kern 1pt} {\kern 1pt} \left[ {{\kern 1pt} S\left( {{\bm{\rho '}} + \frac{1}{2}{\bm{\rho }}} \right) - S\left( {{\bm{\rho '}} - \frac{1}{2}{\bm{\rho }}} \right)} \right] \\
& \delta \left[ {w + \sqrt {{\lambda ^{ - 2}} - {{\left( {{\bm{\rho '}} - \frac{1}{2}{\bm{\rho }}} \right)}^2}}  - \sqrt {{\lambda ^{ - 2}} - {{\left( {{\bm{\rho '}} + \frac{1}{2}{\bm{\rho }}} \right)}^2}} } \right]{d^2}{\bm{\rho '}}
\end{aligned}
\end{equation}
where ${\bm{\zeta }}= \left( {{\bm{\rho }},w} \right)$, $P$ and $P^*$ are the conjugated pair of aperture functions for a circular objective pupil, and $S$ is the intensity distribution of illumination source. The definition of objective pupil function can be expressed as:
\begin{equation}\label{Eq4}
P\left( \bm{\rho} \right) =
\left\{
\begin{aligned}
& 1,\quad \text{if }\left|\bm{\rho}\right|\le {{{\rho }}_{p}} \\
& 0, \quad \text{if }\left|\bm{\rho}\right|>{{{\rho }}_{p}}
\end{aligned}
\right.
\end{equation}
where $\rho_p$ represents the normalized frequency (equals 1) of objective NA. For the most common circular light source, this type illumination is uniformly distributed and the source function is given by:
\begin{equation}\label{Eq5}
S\left( \bm{\rho} \right) =
\left\{
\begin{aligned}
& 1,\quad \text{if }\left|\bm{\rho}\right|\le {{{\rho }}_{s}} \\
& 0, \quad \text{if }\left|\bm{\rho}\right|>{{{\rho }}_{s}}
\end{aligned}
\right.
\end{equation}
where $\rho_s$ denotes the normalized frequency of illumination NA. In this case, the analytic equation of 3D OTF is available in the appendix of \cite{Bao_3DOTF} and \cite{Soto1}.

While there is an oblique coherent source located on the source plane as shown in Fig. \ref{Fig1}, and the distance from this point to the center of pupil is $\rho_s$, written as follows:
\begin{equation}\label{Eq6}
S\left( {u,v} \right) = \delta \left( {u - {\rho _s},v} \right)
\end{equation}
Substituting this source pupil function Eq. (\ref{Eq6}) into Eq. (\ref{Eq3}) results in a 3D POTF for oblique coherent illumination, and the analytical expression can be greatly simplified as:
\begin{equation}\label{Eq7}
\begin{aligned}
{T_p}\left( {u,v,w} \right) = & \frac{{j\lambda }}{{4\pi }}{P^*}\left( {{\rho _s} - u, - v} \right)\delta \left[ {w - \sqrt {{\lambda ^{ - 2}} - \rho _s^2}  + \sqrt {{\lambda ^{ - 2}} - {{\left( {{\rho _s} - u} \right)}^2} - {v^2}} } \right] - \\
& {\kern 1pt} {\kern 1pt} {\kern 1pt} \frac{{j\lambda }}{{4\pi }}{\kern 1pt} {\kern 1pt} {\kern 1pt} P\left( {{\rho _s} + u,v} \right)\delta \left[ {w + \sqrt {{\lambda ^{ - 2}} + \rho _s^2}  - \sqrt {{\lambda ^{ - 2}} - {{\left( {{\rho _s} + u} \right)}^2} - {v^2}} } \right]
\end{aligned}
\end{equation}
where ${P^*}\left( {{\rho _s} - u, - v} \right)$ and $P\left( {{\rho _s} + u,v} \right)$ are two conjugated aperture functions shifted by the oblique coherent source in 3D Fourier space, and these two delta functions are two shifted Ewald spheres defined by functions ${\left( {w + \sqrt {{\lambda ^{ - 2}} + \rho _s^2} } \right)^2}{\rm{ + }}{\left( {{\rho _s} + u} \right)^2}{\rm{ + }}{v^2}{\rm{ = }}{\lambda ^{ - 2}}$ and ${\left( {w - \sqrt {{\lambda ^{ - 2}} - \rho _s^2} } \right)^2} + {\left( {{\rho _s} - u} \right)^2} + {v^2}{\rm{ = }}{\lambda ^{ - 2}}$, respectively. Figure \ref{Fig1} shows the 2D section plots of 3D POTF for three different coherent sources, and these curves are consistent with the previous results in \cite{Soto1,Soto3}. The arc of Ewald sphere is shifted by incident light $S\left( { u,v} \right)$ and limited by the shifted objective pupil functions $P\left( { u,v} \right)$. Besides, the amplitude of POTF is normalized by zero frequency component of AOTF, and the achievable lateral and axial resolution of 3D POTF is extended to the maximum value when both $\rho_s$ and $\rho_p$ are equal to 1.

For arbitrary shape of illumination source, a certain illumination pattern can be discretized into a lot of coherent point sources with finite-size including oblique and upright incident light points. The expression of circular illumination pupil can be written as the sum of delta function in Eq. (\ref{Eq5}). Moreover, the expression for annular illumination aperture can be defined as it follows:
\begin{equation}\label{Eq8}
S({\bm{\rho}}) = \sum\limits_{i = 1}^N {\delta (\bm{\rho} - {{\bm{\rho}}_i})},\quad \left| {{{\bm{\rho}}_i}} \right| \approx {{{{\rho }}_p}}
\end{equation}
where $N$ is the number of all discrete light points on the source plane. Substituting above pupil function Eq. (\ref{Eq5}) or Eq. (\ref{Eq8}) into Eq. (\ref{Eq3}) results in 3D POTF for circular or annular illumination source.

\begin{figure}[!t]
    \centering
    \includegraphics[width=16cm]{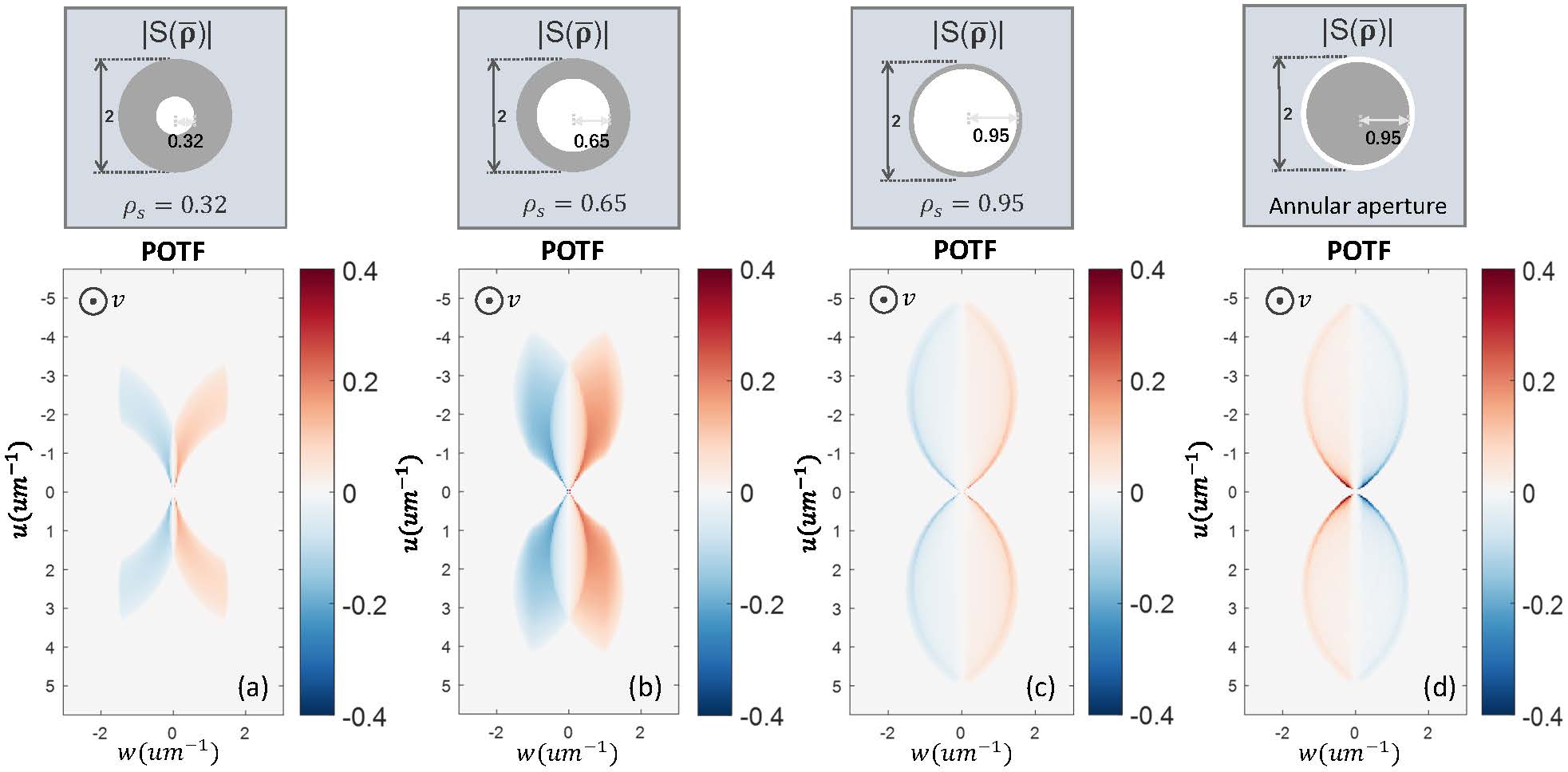}
    \caption{2D plots of 3D POTF section in $u-w$ plane for four different illumination sources including circular and annular illumination aperture with different coherence parameters.}
    \label{Fig2}
\end{figure}

The plots of 3D POTF section in $u-w$ plane for four different illumination sources are illustrated in Fig. \ref{Fig2}. The coherence parameter $\rho_s$ in Figs. \ref{Fig2}(a)-\ref{Fig2}(c) range from 0.32 to 0.95 for circular sources, and the inner radius of annular source is 0.95 in Fig. \ref{Fig2}(d). The lateral and axial resolution are enhanced with the increase of coherence parameter, but the amplitude of POTF is attenuated. These results are coincident with the phenomenon that bigger illumination aperture corresponds to weaker intensity contrast. While the POTF of annular light source not only gives the extension of twice the resolution of coherent diffraction limit, but also improve the low frequency component of POTF around zero frequency point. Thus, the 3D OTFs (including POTF and AOTF) are derived using numerical expression Eq. (\ref{Eq5}) and Eq. (\ref{Eq8}) for arbitrary illumination source, especially for circular and annular shaped illumination sources. Besides, if the asymmetric light source is discretized into the superposition of coherent source like Eq. (\ref{Eq8}), the 3D OTF of DPC model \cite{3D_DPC} can be calculated using the principle above as well.

\section{Principle and implementation of OFC-ODT}

\begin{figure}[!t]
    \centering
     \includegraphics[width=15cm]{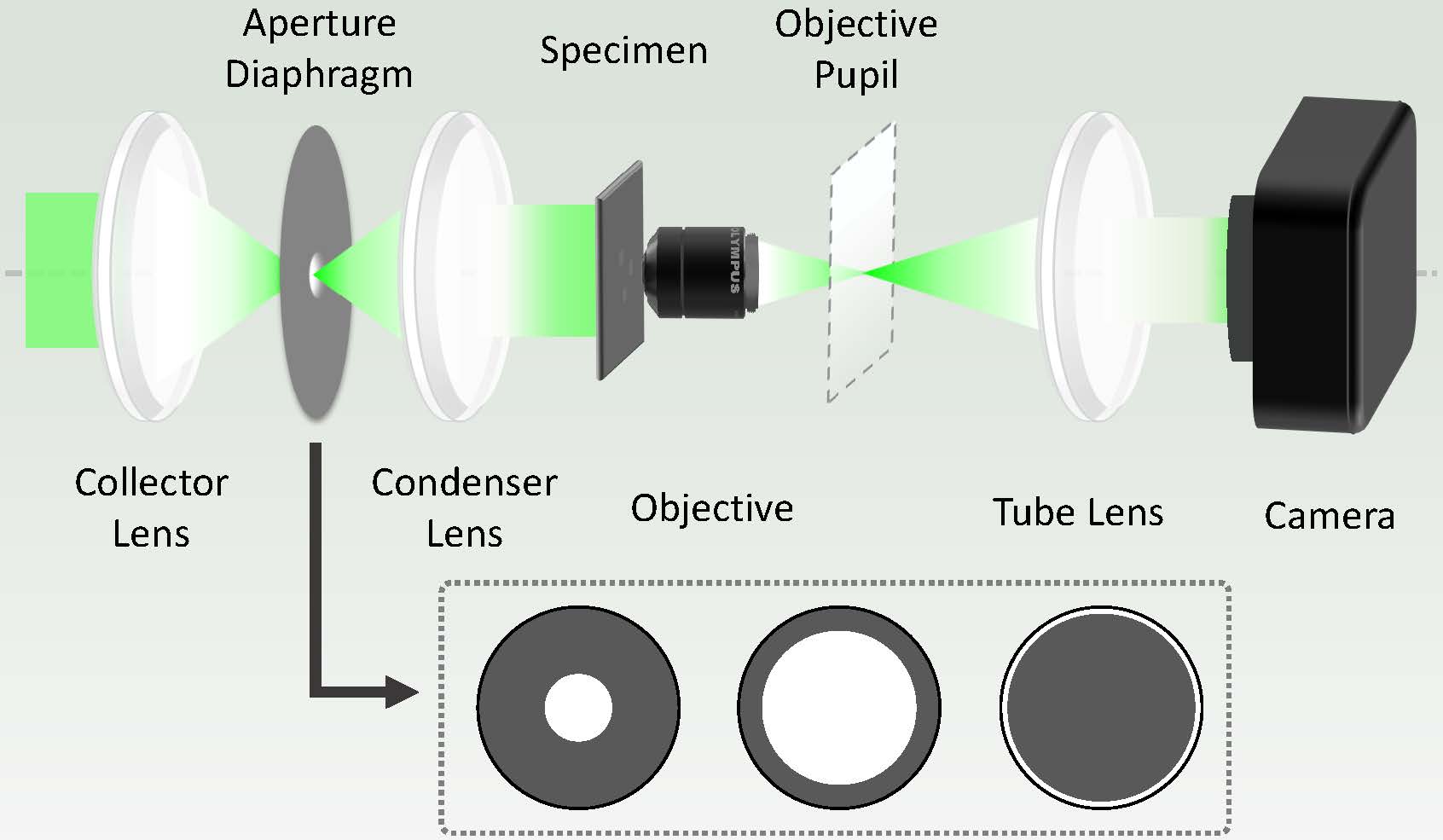}
    \caption{Schematic diagram of experimental setup. The illumination source can be replaced by the circular and annular shaped aperture in the condenser turret and three intensity stacks are captured under respective illumination pattern.}
    \label{Fig3}
\end{figure}

The proposed OFC-ODT technique is implemented based on an inverted commercial microscope (IX83, Olympus) equipped with an oil-immersion objective lens (Olympus UPLSAPO 100 $\times$, NA = 1.4) and an oil-immersion condenser type top lens (Olympus, NA = 1.4), as depicted in Fig. \ref{Fig3}. A halogen white light source with a green interference filter is used for illumination, which can provide quasi-monochromatic lights with narrow bandwidth (central wavelength ${\lambda _0}$ = 550 nm, $\sim$ 10 nm bandwidth). Moreover, the microscope is matched with a scientific CMOS (sCMOS) camera (Hamamatsu, Orca Flash 4.0 V3, 6.5 $\mu$m pixel pitch). The annular aperture used in this work is custom-built by a thin circular glass plate with the opaque regions, and the radius of central opaque circular region, anodized and dyed with a flat-black pigment, is 0.95 (normalized). This annular plate is fitted into an open slot positions in the condenser turret and properly centered in the optical pathway. The axial scanning of intensity images is realized by a motorized focus drive with a step size of 0.1 $\mu$m. In our case, three sets of data are measured under different illumination apertures and each intensity stack contains 100 images with 400 $\times$ 400 pixels. The coherence parameters of two respective circular-shaped apertures are 0.32 and 0.65, and the normalized ring width of annular aperture is 0.05, as shown in Fig. \ref{Fig3}. The spatial sampling rates in $x$, $y$ and $z$ directions are 0.065 $\mu$m, 0.065 $\mu$m and 0.1 $\mu$m, accordingly.

\begin{figure}[!t]
    \centering
    \includegraphics[width=14cm]{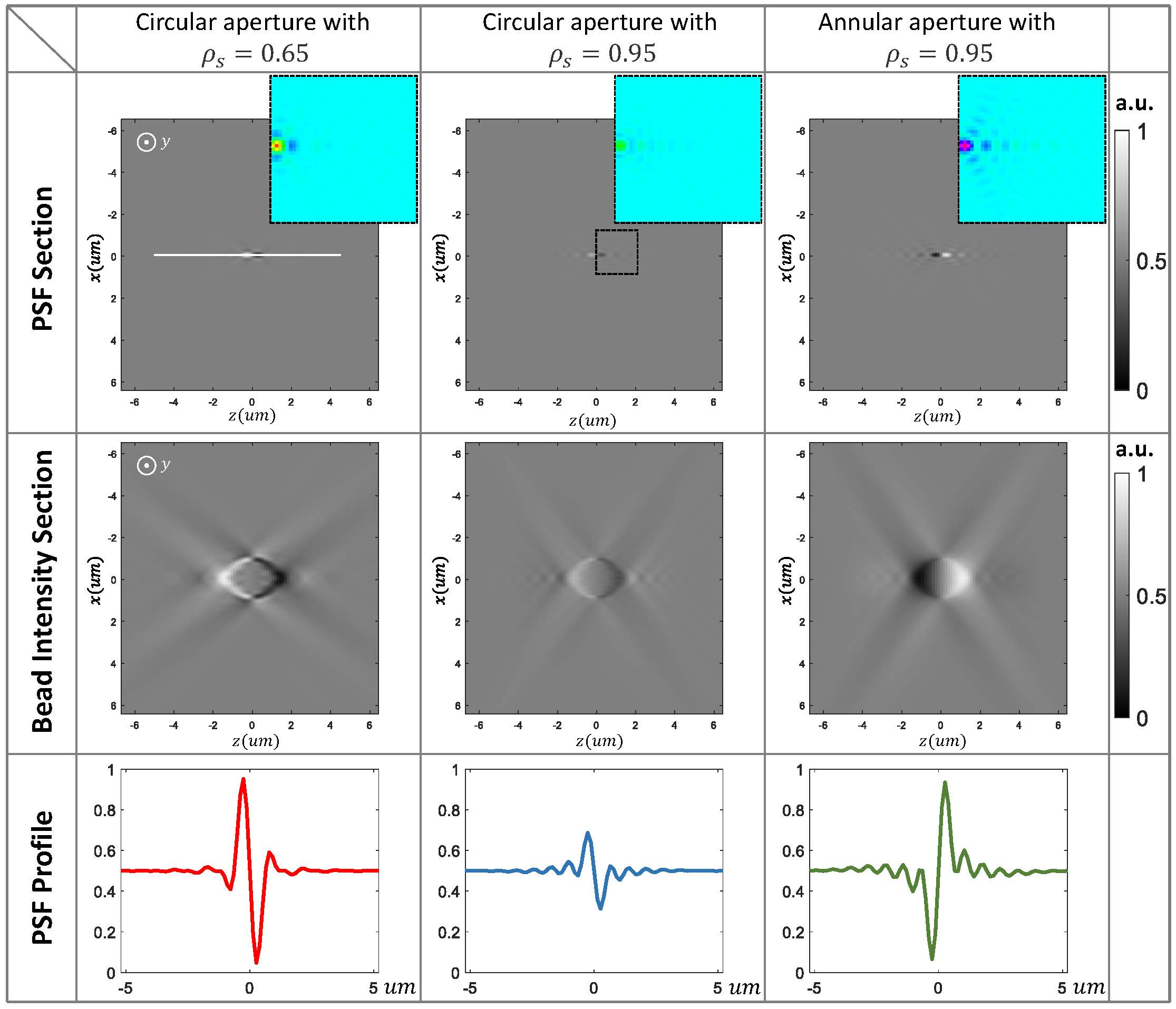}
    \caption{The analysis of PSF for circular apertures with different coherence parameters and annular aperture. The intensity section of an ideal phase micro bead convolved different PFSs and the profiles of PSFs are plotted as well.}
    \label{Fig4}
\end{figure}

Before the detailed description of OFC-ODT method, we will give some analysis and simulation results about the 3D POTF of imaging system using circular and annular illumination apertures. The 3D POTFs for different illumination sources are re-derived in Section 2, and the 2D sections of 3D POTF in $u-w$ plane are plotted in Fig. \ref{Fig2} as well. The first row of Fig. \ref{Fig4} shows the sections of PSF under circular apertures with ${{{\rho }}_s}$ = 0.65, ${{{\rho }}_s}$ = 0.95 and annular aperture with same parameter 0.95. Besides, we zoom in on the selected half subregion of PSF sections and the enlarged HSV color maps show more obvious contrast. The intensity contrast of PSF of circular source with ${{{\rho }}_s}$ = 0.65 and annular one are higher than the PSF contrast of circular source with coherent parameter 0.95. And the diffraction patterns of annular aperture are easier to distinguish. Moreover, an ideal phase micro bead (bead diameter $D_{bead}$ = 2 $\mu$m, bead RI $n_{bead}$ = 1.59, medium RI $n_m$ = 1.58) is convolved with these PSFs and the final bead intensity sections are illustrated in the second row of Fig. \ref{Fig4}. The diffraction angle of defocused intensity images are same with the maximum illumination angle with coherence parameter ${{{\rho }}_s}$ for each aperture. But the contrast of bead intensity image under annular source is opposite to that of the traditional circular source, which is consistent with the OTF and PSF of annular illumination source. The average of these intensity images are normalized to 0.5. In addition, the profiles of PSFs are plotted in the last row of Fig. \ref{Fig4} to give more intuitive results about the characterization of each illumination aperture. Thus, the PSF of annular aperture not only provides strong intensity contrast but also keeps the high frequency components among three illumination apertures.

Next, we provide the quantitative characterizations and simulation results for various illumination apertures. We utilize the direct deconvolution formula (Eq. (4) in \cite{Soto1}) for the 3D RI reconstruction of a simulated 3D phase resolution target. Two same resolution target images are placed in the different axial planes to form a 3D phase object, and the RI contrast between resolution target and surrounding medium is 0.01 with the surrounding medium RI $n_m$ = 1.58. The system parameters are the same with actual experimental setup except that the intensity stack just only contains 128 $\times$ 128 $\times$ 128 pixels and the axial sampling rates is 0.065 $\mu$m. The maximum theoretical lateral resolution ${\lambda _0}/(N{A_{ill}} + N{A_{obj}})$ is about 0.2 $\mu$m, thus the resolution target elements with three pixel periods (0.195 $\mu$m) are chosen as the observed object to be recovered in the lateral direction. As for the axial resolution, the maximum resolution ${\lambda _0}/({{n_m} - \sqrt {n_m^2 - NA_{ill}^2} })$ is about 0.645 $\mu$m, so the distance between two resolution targets is simulated to be 0.65 $\mu$m (10 pixels) in the axial direction.

\begin{figure}[!t]
    \centering
    \includegraphics[width=16cm]{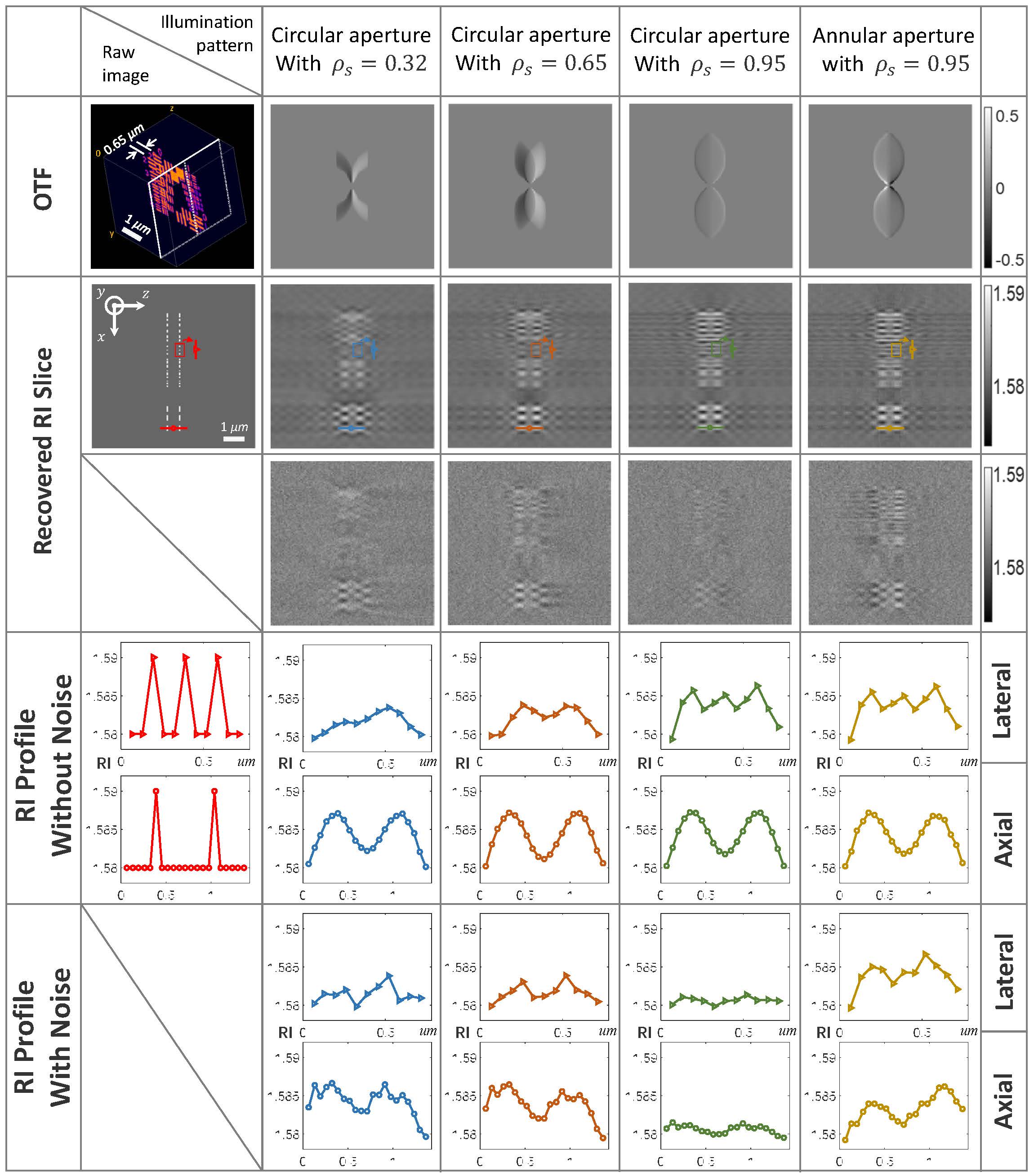}
    \caption{3D RI reconstruction results of a simulated 3D phase resolution target under noise-free situation and the Gaussian noise with a standard deviation of 0.15. The 3D phase object is composed of two same resolution target images placed in the different axial planes, and the distance between two resolution targets is 0.65 $\mu$m in the axial direction. The dimensions of this 3D objet are 128 $\times$ 128 $\times$ 128 pixels with the spatial sample rate 0.065 $\mu$m in both $x$, $y$ and $z$ directions. The RI of resolution target $n$ is 1.58 with the surrounding medium RI $n_m$ = 1.58. Scale bar, 1 $\mu$m.}
    \label{Fig5}
\end{figure}

Figure \ref{Fig5} illustrates the detailed reconstruction results about 3D phase resolution target, and the first column of Fig. \ref{Fig5} are the raw RI slice and RI profiles without noise. It can be seen that the recovered lateral three resolution elements are distinguishable using the circular and annular apertures only both with same coherence parameter 0.95 under noise-free condition. However, the axial RI profiles do not change too much for these four different illumination sources because the lateral resolution improves faster than the axial resolution with the increase of objective NA and illumination NA. In order to characterize the noise sensitivity of different POTFs, we add the Gaussian noise with a standard deviation of 0.15 to intensity images to simulate the noise effect. The recovered results are shown in the last two rows of Fig. \ref{Fig5}, and the lateral RI profile is discernible basically just only under annular illumination aperture. For the corrupted case of lateral and axial profiles, the RI curves of resolution target elements are almost impossible to recover from the intensity image stack due to the poor SNR of its OTF under circular source with ${{{\rho }}_s}$ = 0.95. From the RI Slice reconstruction results of Fig. \ref{Fig5}, the maximum recovered frequency is determined by the sum of objective NA and illumination NA, but the circular shape source with high illumination NA near objective NA degrade the contrast of intensity image and the SNR of 3D OTF. In general, the annular illumination aperture can extend lateral resolution to twice NA of objective and give detectable intensity contrast.

\begin{figure}[!t]
    \centering
    \includegraphics[width=14cm]{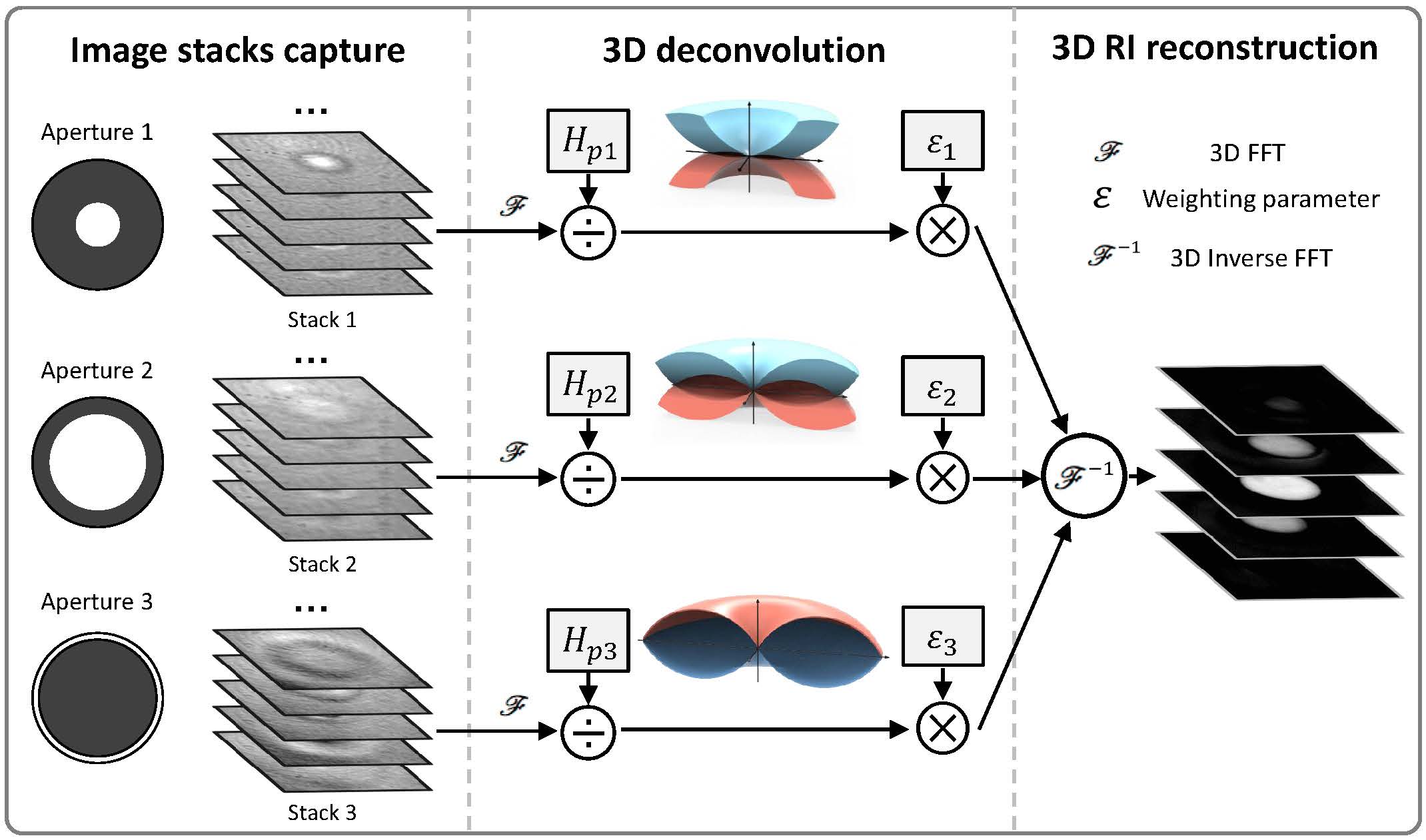}
    \caption{Block diagram representation of the OFC-ODT method.}
    \label{Fig6}
\end{figure}

The 3D POTF of annular aperture can provide robust responses in both low and high frequency on the Ewald sphere, but some other frequency responses in the entire volume transmitted through the incoherent system is not robust enough. So, we make a tradeoff between the quality and speed of 3D diffraction tomographic imaging and employ three intensity stacks for the final RI results. Three intensity stacks are captured under three illumination apertures with different shapes and coherent parameters, and the optimal frequency components of respective 3D POTF are combined together. Figure \ref{Fig6} illustrates a flow chart of OFC method and main idea of our work. Three captured intensity sets are transferred into 3D Fourier spectrum space, and the 3D Fourier spectrum of intensity stack is divided by their own POTF. The 3D scattering potential spectrum sets multiply the different weighting parameters, and the final 3D scattering potential distribution is recovered by implementing an inverse 3D FFT. This process can be written as follows:
\begin{equation}\label{Eq9}
\Phi\left( {\bf{r}} \right) = {{\mathscr{F}}^{ - 1}}\left[ {\frac{{{{\widetilde I }_1}\left( {\bm{\zeta }} \right)}}{{{T_{P1}}\left( {\bm{\zeta }} \right)}}{\varepsilon _1} + \frac{{{{\widetilde I }_2}\left( {\bm{\zeta }} \right)}}{{{T_{P2}}\left( {\bm{\zeta }} \right)}}{\varepsilon _2} + \frac{{{{\widetilde I }_3}\left( {\bm{\zeta }} \right)}}{{{T_{P3}}\left( {\bm{\zeta }} \right)}}{\varepsilon _3}} \right]
\end{equation}
where ${{\mathscr{F}}^{ - 1}}$ denotes the inverse 3D FFT. ${{\widetilde I }\left( {\bm{\zeta }} \right)}$ and ${{T_{P}}\left( {\bm{\zeta }} \right)}$ are the 3D intensity Fourier spectrum and POTF of a certain aperture, respectively. And ${\varepsilon}$ is the weighting parameter for each Fourier spectrum set and calculated by linear least-squares method among three sets of POTF:
\begin{equation}\label{Eq10}
{\varepsilon _i} = \frac{{T_{_{Pi}}^*\left( {\bm{\zeta }} \right){T_{_{Pi}}}\left( {\bm{\zeta }} \right)}}{{{{\left| {{T_{P1}}\left( {\bm{\zeta }} \right)} \right|}^2} + {{\left| {{T_{P2}}\left( {\bm{\zeta }} \right)} \right|}^2} + {{\left| {{T_{P3}}\left( {\bm{\zeta }} \right)} \right|}^2}}}.
\end{equation}
Substituting Eq. (\ref{Eq10}) into Eq. (\ref{Eq9}) results the final expression of 3D RI reconstruction:
\begin{equation}\label{Eq11}
\Phi\left( {\bf{r}} \right) = {{\mathscr{F}}^{ - 1}}\left[ {\frac{{{{\widetilde I }_1}\left( {\bm{\zeta }} \right)T_{_{P1}}^*\left( {\bm{\zeta }} \right){\rm{ + }}{{\widetilde I }_2}\left( {\bm{\zeta }} \right)T_{_{P2}}^*\left( {\bm{\zeta }} \right){\rm{ + }}{{\widetilde I }_3}\left( {\bm{\zeta }} \right)T_{_{P3}}^*\left( {\bm{\zeta }} \right)}}{{{{\left| {{T_{P1}}\left( {\bm{\zeta }} \right)} \right|}^2} + {{\left| {{T_{P2}}\left( {\bm{\zeta }} \right)} \right|}^2} + {{\left| {{T_{P3}}\left( {\bm{\zeta }} \right)} \right|}^2} + \alpha }}} \right]
\end{equation}
where $\alpha$ is regularization parameter according to the noise level of intensity images. Although this type idea of OFC has been proposed in the 2D phase imaging methods \cite{OFC1,OFC2}, we extend this ideas to the 3D tomographic imaging, especially for the transfer function between different illumination modalities. In addition, the algebraic iterative reconstruction algorithms with non-negative and total variation regularization \cite{TOMO_TV1,TOMO_TV2} are applied to the final RI stack and the missing cone effect is alleviated for more accurate 3D reconstruction results.

As for the image acquisition and data processing time, the exposure times of two circular apertures and annular aperture are 15 ms, 10 ms and 30 ms in experiments, respectively. All the experimental data is processed by Matlab software (MATLAB R2016a) and a personal computer (Intel Core i7-8700K, 3.7 GHz, 16 GB DDR4 RAM), and the time required for all computation processes (including two times iterative constraint) of intensity stacks field of view (FOV) (400 $\times$ 400 $\times$ 100 $\times$ 3) is about 4.5 seconds. If we want to make the reconstruction process of a region of interest faster, only one captured intensity stack under annular illumination with an exposure time of 2 ms should be used, and it is possible to achieve a rate of about 5 3D frames per second for live cell imaging further.

\section{Experimental results}

\begin{figure}[!b]
    \centering
    \includegraphics[width=16cm]{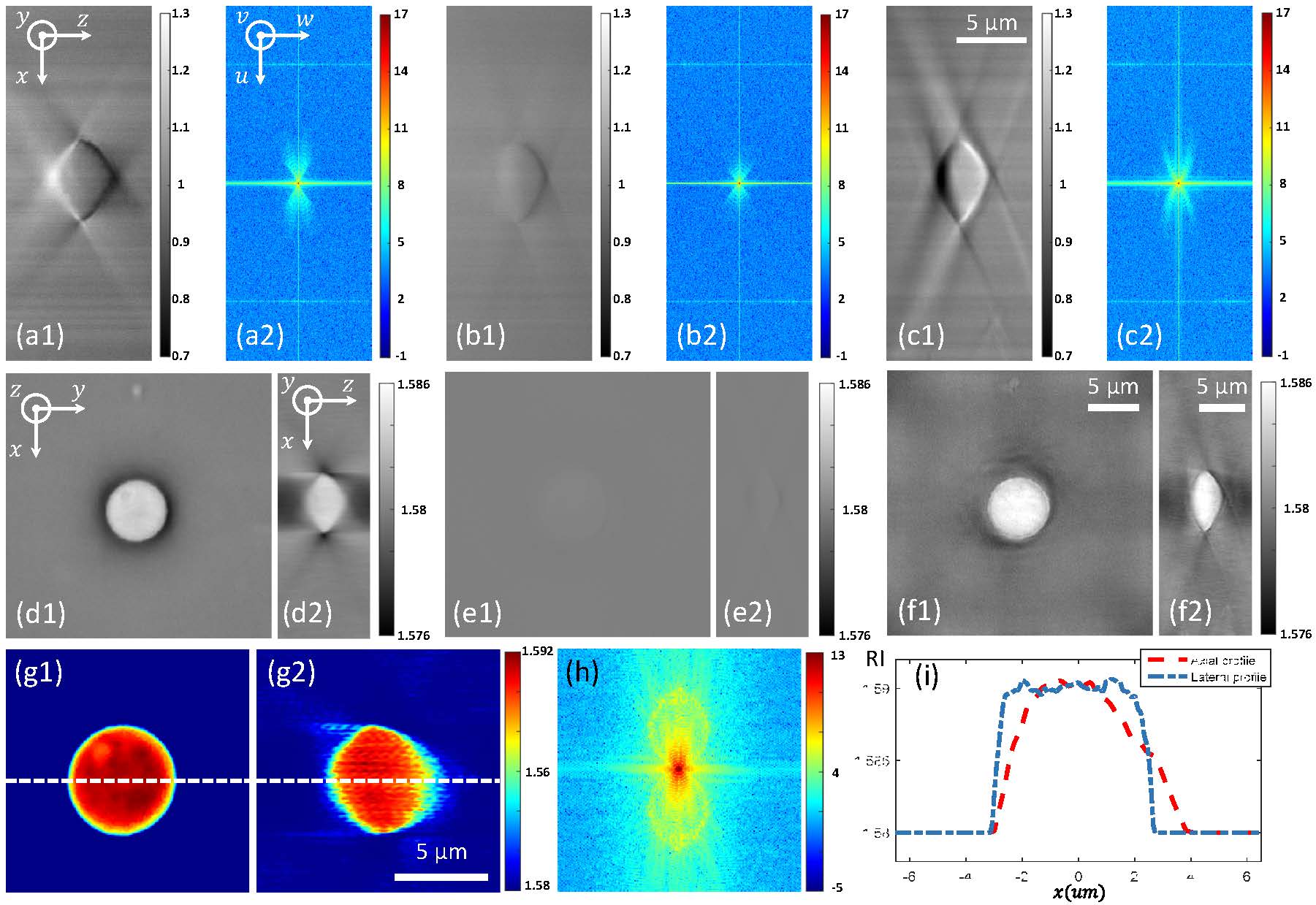}
    \caption{RI experimental results of micro polystyrene bead with 6 $\mu$m diameter. (a-c) Raw images of captured intensity stacks and Fourier spectrum sections under three different illumination apertures. (d-f) Recovered lateral and axial slices using the direct deconvolution equation. (g) Final 3D RI slices with same pixel sampling in all directions. (h) Final recovery of Fourier spectrum after iterative constraint. (i)  Axial and lateral RI profiles of reconstructed micro bead. Scale bar, 5 $\mu$m.}
    \label{Fig7}
\end{figure}

First, we implement the OFC-ODT technique to a control sample for the quantitative 3D RI reconstruction. The micro polystyrene bead (Polysciences, $n$ =1.59 at ${\lambda _0}$ = 589 nm) with 6 $\mu$m diameter is immersed in oil (Cargille, $n_m$ =1.58), and the sCMOS camera, oil-immersion objective lens and condenser lens (maximum NA = 1.4 both) are employed for the capture of intensity stacks. Sampling frequencies for the following experiments are 0.065 $\mu$m$^{ - 1}$, 0.065 $\mu$m$^{ - 1}$ and 0.1 $\mu$m$^{ - 1}$ in $u$, $v$ and $w$ direction in Fourier space, accordingly. Figure \ref{Fig7} displays the comparative results between three type apertures and the final results of micro sphere using OFC-ODT method. In Figs. \ref{Fig7}(a)-\ref{Fig7}(c) the images present the axial raw intensity and Fourier spectrum slices for circular illumination aperture with 0.9 NA ($\rho_s$ = 0.65), 1.33 NA ($\rho_s$ = 0.95) and annular aperture with 1.4 NA ($\rho_s$ of outer ring equals 1). The average values of three sets of intensity slices are normalized to 1, and the display dynamic range of intensity spectrum is fixed and set from -1 to 17. It can be seen that the image contrast of Figs. \ref{Fig7}(a1) and \ref{Fig7}(c1) is higher than Fig. \ref{Fig7}(b1), which is consistent with the simulation results illustrated in Fig. \ref{Fig4}. The transmitted Fourier spectrum of intensity stack is same as their own POTF in shape and amplitude additionally.

Figures \ref{Fig7}(d)-\ref{Fig7}(f) are the recovered lateral and axial slices using the direct deconvolution equation, and the regularization parameter $\alpha$ is set to ${10^{ - 3}}$ for OFC method and $5 \times {10^{ - 5}}$ for Fig. \ref{Fig7}(e), respectively. Although the recovered axial resolution of RI slice under fully opened circular aperture in Fig. \ref{Fig7}(e2) are better than the result in Fig. \ref{Fig7}(d2), the RI contrast in Fig. \ref{Fig7}(e) is too weak and the 3D POTF of fully opened circular aperture gives poor response in pass-band of the whole transmitted volume. After two times of iterative constraint, the final RI distributions and Fourier spectrum slices of reconstructed micro bead are presented in Figs. \ref{Fig7}(g) and \ref{Fig7}(h) with same sampling rate of 0.065 $\mu$m in all directions. From the spectrum image of micro bead RI, four semicircular arcs form the entire frequency volume transmitted through the incoherent system, and the axial and lateral imaging resolutions are extended to maximum theoretical limits. However, the recovered bead suffers from some elongation and attenuation along the axial direction due to the alleviated missing cone issue, see Fig. \ref{Fig7}(i). Overall, the results of micro polystyrene bead validate the success and accuracy of proposed OFC-ODT method basically.

\begin{figure}[!b]
    \centering
    \includegraphics[width=15cm]{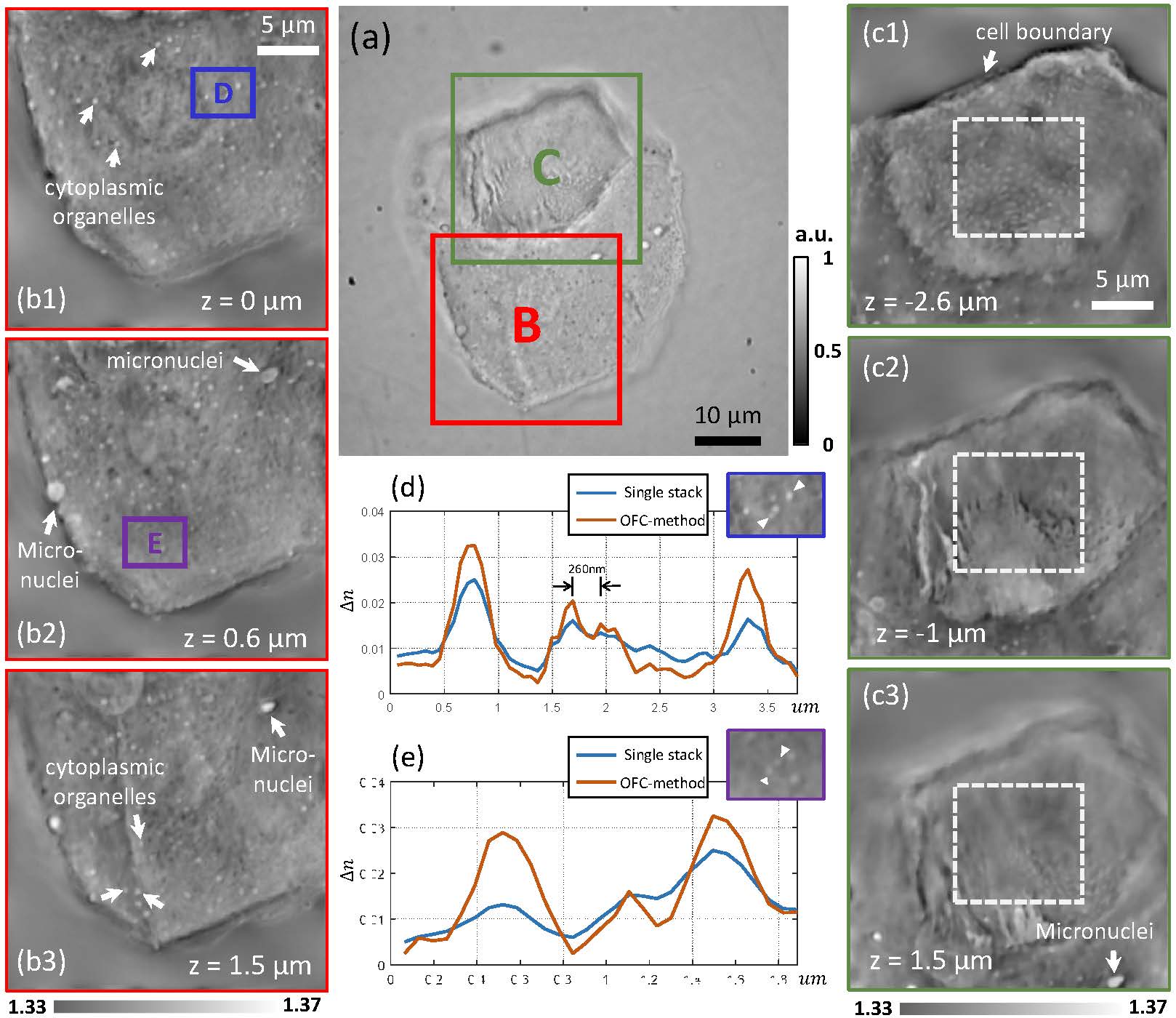}
    \caption{Tomographic reconstruction of human buccal epithelial cell. (a) The captured intensity image of full sensor FOV under circular illumination aperture with $\rho_s$ = 0.65. (b-c) Detailed RI slices at different axial planes of two enlarged regions, see also \textcolor{blue}{Visualization 1}. (d-e) Comparative profile lines of quantitative RI measurement between single stack and OFC method for two selected small regions. The achievable lateral resolution of OFC-ODT technique up to 260 nm. Scale bar denotes 10 $\mu$m and 5 $\mu$m, respectively.}
    \label{Fig8}
\end{figure}

Then, in order to test the performance of our technique in its intended biomedical application, the unstained fresh human buccal epithelial cell (cheek cell) smear is used for the 3D RI measurement. The cheek cells are placed in the gap between two thin $no$. 0 coverslips with RI matching 0.9\% sodium chloride solution (${n_{water}}$ = 1.33). Figure \ref{Fig8}(a) is the captured intensity image of full sensor FOV under circular illumination aperture with $\rho_s$ = 0.65. Two representative regions, labeled B and C, are selected for the detailed analysis, and the enlarged RI reconstruction slices of these regions at three different axial positions are illustrated in Figs. \ref{Fig8}(b) and \ref{Fig8}(c), respectively. There, the optically micronuclei and some large cytoplasmic organelles are shown with distinguishable RI contrast and clarity in Figs. \ref{Fig8}(b1)-\ref{Fig8}(b3). In Fig. \ref{Fig8}(c), the region C shows the cell boundary, bacteria (diplococci mostly) on surface of cells and the RI changing of the squamous structures in cytoplasm from the top to the ground plane. Besides, two measured comparative RI line profiles of two sub-region in Fig. \ref{Fig8}(b) using several stacks and single stack ($\rho_s$ = 0.65) of intensity are presented in Figs. \ref{Fig8}(d)-\ref{Fig8}(e). These plot lines demonstrate that the significant improvement of high frequency features using OFC-ODT compared to conventional PC-ODT method. The smallest yet clearly visible cytoplasmic organelle structures have a separation of only 260 nm, even though it is not possible to identify mitochondria with the light microscope. The final recovered stacks of RI images along with axial direction for two different regions are provided in \textcolor{blue}{Visualization 1}.

\begin{figure}[!t]
    \centering
    \includegraphics[width=15cm]{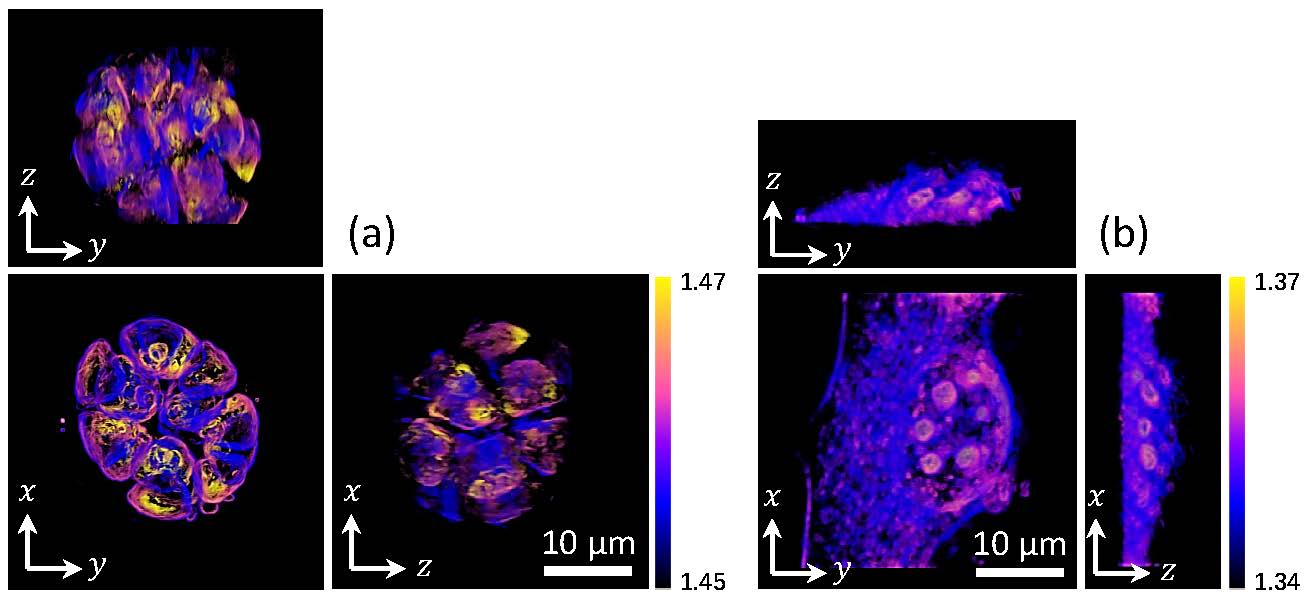}
    \caption{3D RI tomograms rendering of (a) Pandorina morum algae and (b) HeLa cell in $x-y$, $z-y$ and $x-z$ planes. Scale bar, 10 $\mu$m.}
    \label{Fig9}
\end{figure}

Finally, as evidenced above, the proposed OFC-ODT technique can provide an optical RI sectioning capability and is well suited for 3D quantitative imaging of biological samples. This diffraction tomographic system is applied to a bleached paraffin section of Pandorina morum algae ($P. morum$) and a culture-medium immersed HeLa cell further. $P. morum$ is a genus of green algae composed of 8, 16, or sometimes 32 cells and these cells are held together at their bases to form a sack globular colony surrounded by mucilage. And the cells of $P. morum$ form an ellipsoid spheroid with a distinct anterior-posterior polarity. Moreover, the $P. morum$ is cleaned with ${H_2}{O_2}$ to remove the highly absorbing algae and immersed in the fixed embedding medium (paraffin wax, ${n_{wax}}$ = 1.45). The RI of culture medium ${n_{m}}$ of HeLa cells is 1.34. Figure \ref{Fig9} shows the 3D rending results of $P. morum$ and HeLa cell in $x-y$, $z-y$ and $x-z$ directions, respectively. The nucleolus inside cell have significant RI contrast compared to the surrounding cytoplasm and these rendered RI images give vivid reconstruction results. Two video files \textcolor{blue}{Visualization 2} and \textcolor{blue}{Visualization 3} present the 3D animated version and reveal much clearer information about the actual structure of the specimen.

\section{Conclusion and discussion}

In summary, we demonstrate a novel tomographic technique termed OFC-ODT by combining the optimal frequency components of three transfer function sets together in a traditional bright-field transmission microscope with different illumination apertures. The expression of 3D OTF for arbitrary illumination source, especially for annular illumination aperture or asymmetric case, is re-derived. The inversion of 3D POTFs are combined by utilizing linear least-squares method and applied to three intensity stacks. Besides, the characterizations  and imaging performance of annular illumination source are thoroughly explored and the quantitative analysis and simulation results based on a 3D phase resolution target are proposed as well. The 3D POTF of annular aperture can achieve twice the NA of coherent systems and provide robust responses in both low and high frequency on the Ewald sphere. Finally, the 3D RI reconstruction and rendering visualization of control micro sphere and various biological specimens validate the success of this method.

In order to ensure the coherence and the maximum illumination NA of the annular source, the width of nonopaque ring should be designed to be between 5\% and 10\% of the outer circle of the annulus. Due to the limited light efficiency, the total brightness of the annular illumination may not be enough to generate intensity images with reasonably good SNR. And the recording of several stacks of intensity images is time-consuming for single stack method. Despite the drawback presented above, we can use only one set of intensity stack under annular aperture with  higher quantum efficiency camera and brighter light source. But the most important revelation of this work is to provide a potential way for 2D and 3D phase imaging by using the modulation of illumination or programmable aperture and make these phase imaging method compatible with conventional bright-field microscopes.

\section*{ACKNOWLEDGMENTS}
This work was supported by the National Natural Science Fund of China (61722506, 61505081, 111574152), Final Assembly `13th Five-Year Plan' Advanced Research Project of China (30102070102), National Defense Science and Technology Foundation of China (0106173), Outstanding Youth Foundation of Jiangsu Province of China (BK20170034), The Key Research and Development Program of Jiangsu Province, China (BE2017162), `Six Talent Peaks' project of Jiangsu Province, China (2015-DZXX-009), `333 Engineering' Research Project of Jiangsu Province, China (BRA2016407), Fundamental Research Funds for the Central Universities (30917011204, 30916011322), Open Research Fund of Jiangsu Key Laboratory of Spectral Imaging \& Intelligent Sense (3091601410414).

\section*{DISCLOSURES}
The authors declare that there are no conflicts of interest related to this article.

\end{document}